\documentclass[11pt]{article}

\usepackage{fancyhdr}

\def\eqalign#1{\null\,\vcenter{\openup\jot
  \ialign{\strut\hfil$\displaystyle{##}$&$\displaystyle{{}##}$\hfil
      \crcr#1\crcr}}\,}

\usepackage{color}
\usepackage{hyperref}
\def\texorpdfstring#1{}
\definecolor{iblue}{RGB}{65,105,225}
\definecolor{ired}{RGB}{220,20,60}
\definecolor{igreen}{RGB}{50,205,50}
\definecolor{ipurple}{RGB}{75,0,130}
\definecolor{iochre}{RGB}{218,165,32}
\definecolor{iteal}{RGB}{51,204,204}
\definecolor{imauve}{RGB}{204,51,153}

\usepackage{graphicx}
\usepackage{amsfonts}
\usepackage{setspace}

\def\iniz{\setcounter{equation}{0}{%
\rhead{\thepage}\lhead{{{{\small\bf\thesection:}
\small \ \SEC\ \  \tiny\today}}}}}

\let\a=\alpha     \let\d=\delta 
      \let\k=\kappa 
\let\m=\mu    \let\n=\nu         \let\p=\pi    
\let\s=\sigma \let\t=\tau    
     
 \let\D=\Delta   
         \let\Ps=\Psi

\def\V#1{{\bf#1}}
\def\*{\vskip 3mm}\def\0{\noindent}
\def\be{\begin{equation}}
\def\ee{\end{equation}}
\def\bea{\begin{eqnarray}}
\def\eea{\end{eqnarray}}

\def\xx{{\V x}}\def\kk{{\V k}}
\def\CC{{\cal C}}

\let\dpr=\partial\let\fra=\frac
\def\EE{{\cal E}}\def\DD{{\cal D}}

\def\TT{{\mathcal T}}

\def\Eq#1{\label{#1}}
\def\equ#1{(\ref{#1})}
\def\lis#1{\overline#1}
\def\defi{{\buildrel def\over=}}
\def\media#1{{\langle\,#1\,\rangle}}
\def\bra#1{{\langle#1|}}\def\ket#1{{|#1\rangle}}

\def\otto{\,{\kern-1.truept\leftarrow\kern-5.truept\to\kern-1.truept}\,}

\def\tende#1{\,\vtop{\ialign{##\crcr\rightarrowfill\crcr
 \noalign{\kern-1pt\nointerlineskip} \hskip3.pt${\scriptstyle
 #1}$\hskip3.pt\crcr}}\,}
\def\ie{{\it i.e.\ }}
\def\eg{{\it e.g.\ }}

\def\Eq#1{{\label{#1}}%
}

\newdimen\xshift \newdimen\xwidth \newdimen\yshift \newdimen\ywidth
\def\ins#1#2#3{\vbox to0pt{\kern-#2\hbox{\kern#1 #3}\vss}\nointerlineskip}

\def\eqfig#1#2#3#4#5{
\par\xwidth=#1 \xshift=\hsize \advance\xshift
by-\xwidth \divide\xshift by 2
\yshift=#2 \divide\yshift by 2%
{\hglue\xshift \vbox to #2{\vfil
#3 \includegraphics{#4.eps}
}\hfill\raise\yshift\hbox{#5}}}
\DeclareGraphicsExtensions{.eps,.png,.pdf}

\def\BDpr {\mbox{\boldmath$ \partial$}}

\def\({\left(}
\def\){\right)}

\def\V#1{{\vec #1}}
\def\uu{{\bf u}}
\def\vv{{\bf v}}

\def\gg{{\bf g}}
\def\kk{{\bf k}}
\def\xx{{\bf x}}
\def\lis#1{{\overline#1}}

\def\alert#1{{\color{ired}#1}}
\def\alertb#1{{\color{iblue}#1}}

\title{\bf Finite thermostats in classical and quantum nonequilibrium}
\begin{document}

\maketitle

\centerline{Giovanni Gallavotti}
\centerline{INFN-Roma1 and Rutgers University\footnote{
    P.le Aldo Moro 2, 00185 Roma, Italy\\
    \hglue7mm email: giovanni.gallavotti@roma1.infn.it\\
    \hglue7mm homepage \tt http://ipparco.roma1.infn.it/$\sim$giovanni}}

\begin{abstract}
  Models for studying systems in stationary states but out of equilibrium
  have often empirical nature and very often break the fundamental time
  reversal symmetry. Here a formal interpretation will be discussed of the
  widespread idea that, in any event, the particular friction model choice
  should not matter physically.  The proposal is, quite generally, that for
  the same physical system a time reversible model should be
  possible. Examples about the Navier-Stokes equations are given.
\end{abstract}
\0{\sl keywords:} {\it\small Friction, Reversibility, Irreversibility,
  Fluctuation Theorem, Nonequilibrium Ensembles, Quantum Dissipation.}

\def\SEC{Thermostats}
\section{\SEC}
\label{sec1}
\iniz

A mechanical system, described by coordinates $x\in R^N$, moving subject to
an equation $\dot x=F(x)$ is ``$I$-time reversal symmetric'' if there is a
coordinate transfomation $x\to Ix$ with $I^2=1$ and such that if $x(t)\defi
S_{t} x$ is a motion then also $Ix(-t)$ is a motion, {\it i.e.} $I S_{t}\equiv
S_{-t} I$.

A mechanical system is said {\it subject to thermostat forces} if the
equations of motion have the form $\dot x=h(x)+f(x)-L(x)$ where $x\in R^N,
\,N\le\infty$, $h(x)$ is a force which in absence of the other two terms
would admit a conserved energy (typically
$\dot x=h(x)$ is a Hamiltonian system), $f(x)$ is a ``driving force''
capable of performing work thus changing the energy and the equation $\dot
x=h(x)+f(x)\equiv F(x)$ admits a time reversal symmetry $I$ in the above sense;
$L(x)$ is a dissipative force that will keep at bay the energy increase due
to the work of the force $f$ so that the motion may possibly attain a
stationary (nonequilibrum) state.

Systems subject to thermostats are obtained either by introducing
empirically a dissipating force characterized in terms of few
parameters (``friction coefficients'' or ``transport coefficients'') into
the equations of motion (which may be either ode's or pde's) or by
imagining the system in interaction with one or more ``external systems''
of particles, extending to infinity and asymptotically in thermal
equilibrium, as envisaged in \cite{FV963}. Or, also, by introducing
empirically forces which in some way absorb in average the work done by the
external forces.

For instance\\
(1) In the simple electric resistor model, in which $N$ point particles
move in a periodic box elastically colliding with fixed spherical obstacles
and are subject either to a constant uniform field $E$ (hence non conservative)
and to a friction $-\n \dot x_j$, which is the empirical thermostat force
(forbidding energy blow up but enforcing breaking of time reversal for
$I(x_j,\dot x_j)=(x_j,-\dot x_j$) or, in
``Drude's model'', \cite{Be964}, rescaling the kinetic energy of a particle
to a fixed value $\frac32 k_BT$ at each collision.
\\
(2) In the ``Lorenz96'' model
\be \dot x_j=\ x_{j-1}(x_{j+1}-x_{j-2})+ F - \n  x_j,\qquad \n>0,\ j=1,\ldots,N
\Eq{e1.1}\ee
with $x_{N+1}=x_1$, which at $F=0,\n=0$ conserves $E=\sum_i x_i^2$, the
(thermostat force is $-\n x_j$, breaking the time reversal $Ix_j=-x_j$) .
\\
(3) A further example, discussed in more detail below, is provided by
the Navier-Stokes equations for an incompressible velocity field
$\vv(\xx,t)$, called here ``irreversible Navier-Stokes equation'' or
``I-NS'':
\be{\dpr_t\, {\vv}=-{(\V v\cdot\V \dpr)}\, {\vv}+\n
  \,\D \,{\vv} +F\,{\gg}-{\BDpr} p'}, \quad \BDpr\cdot\vv=0, \qquad
   \hbox{[I-NS]}\Eq{e1.2}\ee
in dimension $2$ and considered, for simplicity, in a periodic container of
side $L$, under a forcing of strength $F$ with fixed ``structure'' $\gg$,
pressure $p'$ and viscosity $\n$. This is an example (a pde in which
$N=\infty$): here the artificial thermostatting force is represented by the
viscosity stress $-\n \D\vv$. At $\n=0,F\ge0$ the flow of Eq.\equ{e1.2} is
time reversible, with time reversal $I$ defined by $I\vv=-\vv$, and
conserves $E=\int \vv^2(\xx) d\xx$ as well as $E_n=\int
(\V\dpr\vv(\xx))^2d\xx$ (because dimension is $2$): but the flow breaks
time reversal and both conservation laws if $\n\ne0$.  \\
(4) Modeling dissipation is difficult in studying nonequilibrium states of
quantum systems because the dissipative forces lead to modify the
Schr\"odin\-ger equation of the system by the addition of non conservative
forces: not a simple task since they just cannot be represented by self
adjoint operators. Restricting the analysis to heat conduction problems,
the simplest thermostats to consider are the infinite ones as proposed in
\cite{FV963}:

\hglue4cm\includegraphics[width=80pt, height=80pt]{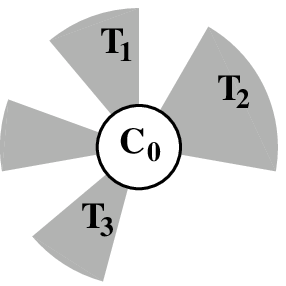}

\begin{spacing}{0.2}
\0\tiny Fig.1: A finite system $\CC_0$,
called ``test system'', interacts with one or more ``interactions sytems''
supposed infinite and asymptotically in thermal equilibrium at temperatures
$T_j$. All systems are quantum particles systems, \cite{FV963}.

\end{spacing}

\vglue5mm These systems are difficult to study theoretically unless the
particles constituting the thermostats are free gases, \cite{FV963}. And
simulations are, otherwise, impossible due to the infinite size of the
thermostats: infinite thermostats (either classical or quantum) cannot be
simulated without approximating them by finite systems.  Therefore models
have been developed to build thermostat models suitable for simulations of
quantum systems, hence either exactly soluble (a very rare case, as in
\cite{FV963,ABGM971,ABGM972}) or finite dimensional, \eg 
\cite{MM978,MCT993,ACCCEF011}.  \*

In many cases several thermostat models have been introduced, or may be
introduced, and the proposal here is to formulate a
unification and an interpretation of the diverse theories as a
generalization of the theory of ensembles of equilibrium statistical
mechanics. The idea will be illustrated in detail through the examples of
the irreversible 2D Navier-Stokes equation, ``I-NS'', and the Erehenfest
model of quantum forced system, \cite{MM978}.

Since the founding fathers time, microscopic reversibility has been the
starting point of any determinstic description (either microscopic or
macroscopic) of the evolution of systems. However in more recent times
progress has been achieved in the analysis of stochastic models of
evolution, classical or quantum, with attention to the guiding idea
that thermal noise must at all times obey the classical or quantum version
of the fluctuation-dissipation theorem; and it has been observed that this
implies strong constraints in models of dissipative evolutions,
\cite{HT982,HI005}: it is proposed that the familiar consequences of the
microscopic time-reversal symmetry must hold also in the approximate stochastic
models for the evolutions.

In the present work thermal noise is not considered: the idea is that a
large class of deterministic models can be equivalently described by
different (macroscopic or not) equations whose stationary states attribute
(at least in stronly chaotic regimes) the same averages, as well as same
statistical fluctuations, to large classes of observables: chaotic motion
is implicitly regarded as sounce of noise. And the equivalence has
interesting relations with the theory of equivalent ensembles in
equilibrium statistical mechanics, at least at strong chaos, thus making
possible testing the fluctuation relation, \ie an extension of the
fluctuation-dissipation relation beyond the perturbative
linear regime, \cite{GC995,Ga996a,Ga996b,Ga998}, via experiments on systems in
which friction plays a role but is usually modeled via phenomenological
constants.

\def\SEC{Nonequilibrium ensembles}
\section{\SEC}
\label{sec2}
\iniz

Consider the I-NS equation Eq.\equ{e1.2} and cast it in dimensionless form by
setting $\vv(\xx,t)=V \uu(\xx/L,t/T)$ with the rescaling parameters
$V,T,L$ so that $L=1$, $\fra{TV}{L}=1$ and $\fra{FT}V=1$. Introducing the
Reynolds number $R=(\fra{F L^3}{\n^2})^{\fra12}$ the dimensionless equation
for $\uu(\xx,t)$ results:
\be{\dpr_t\, {\uu}=-{(\V u\cdot\V \dpr)}\, {\uu}+\frac1R
  \,\D \,{\uu} +{\gg}-{\BDpr} p}, \quad \BDpr\cdot\uu=0, \qquad
   \hbox{[I-NS]} \Eq{e2.1}\ee
Fixed $\gg$, for each $R$ the flow generated by the above I-NS equation
will define a stationary state which will be supposed, for the time being,
unique for $R$ large enough: the state will be described by a stationary
probability distribution, called here $\m^{mc}_R(d\uu)$, on the 
(smooth) velocity fields $\uu$.
\footnote{
\tiny
For $R$ small there might exist several attractors, \cite{FT985}, and the
stationary states for the flow might be non unique so that more parameters
would be needed to indicate the stationary states of the I-NS flow,
\cite{FT985}. But this should not happen for large $R$ (identifying states
that differ by symmetries of the equation, \cite{Fr983}). See below for
more general cases, like small $R$.
}

Hence for each $R$ the system will reach a stationary state, which will be
supposed unique {\it for simplicity} and will be
represented by an invariant probability distribution of the velocity fields
a PDF denoted $\m^c_R(d\uu)$.

As $R$ varies a collection $\EE^c$ of probability distributions is obtained
which will be called the {\it ``viscosity ensemble''} for
I-NS.\footnote{\tiny Recall that after the rescaling the $\frac1R$ is
  identified with the viscosity $\n$.}

The I-NS equation is a macroscopic model of a microscopic system of
interacting particles following Newton's equations subject to the condition
that, by some unspecified mechanism modeled by the stress $\n\D \uu$, the
work done by the external force $\gg$ is dissipated (in average).

The same effect can be achieved in other ways: for instance consider the
equation for the velocity field, conveniently represented via the Fourier's
transform as
$\uu(\xx)=\sum_{\kk} \uu_\kk e^{2\p i \kk\cdot\xx}, \,\kk\cdot\uu_\kk=0$
with $\kk=(k_1,k_2)$ integers:

\be\eqalign{
&
\dot{\uu}+(\V u\cdot \V\dpr)\uu=-\BDpr p+ \gg+\a(\uu)\D\uu,\qquad
\BDpr \cdot\uu=0, \qquad\hbox{\rm [R-NS]}
\cr
&\a(\uu)\defi
\frac{\int_0^{2\p} \gg(\xx)\cdot(\D \uu)(\xx) d\xx}{\int_0^{2\p} (\D
    \uu(\xx)^2) d\xx}\equiv
  \frac{\sum_{\kk} \kk^2\,\gg_{\kk}\cdot \uu_{-\kk}}{\sum_{\kk}
    \kk^4|\uu|^2},\qquad \kk\cdot\uu_\kk=0
  \cr
}\Eq{e2.2}\ee
in which $\a$ is so defined that the ``\alert{dissipation}'' observable
$\DD(\uu)=\int (\V\dpr \uu(x))^2 dx$ is an \alertb{exact constant of
  motion}. Eq.\equ{e2.2} will be called ``R-NS'', for {\it ``reversible
Navier-Stokes''}, because the transformaion $I\uu(x)\defi -\uu(x)$ is a
time reversal for the R-NS equation above (\ie it $\uu(x,t)$ is a solution
also $-\uu(x,-t)$ is a solution).

Therefore the stationary states of the R-NS will be parameterized by the
value $En$ of $\DD(\uu)$: it will be a PDF $\m_{En}^{mc}(d\uu)$ on the
velocity fields which will be supposed (for the time being) unique for $En$
large.

As $En$ varies the distributions $\m_{En}^{mc}(d\uu)$ form a family of PDF's
that will be denoted $\EE^{mc}$, and will be called the {\it reversible
  viscosity ensemble}.

Remark that the multipler $\a(\uu)$ is defined no matter which flow,
Eq.\equ{e2.1} or Eq.\equ{e2.2}, is considered. In the flows defined by the
I-NS or R-NS $\a(\uu)$ is a fluctuating variable, and in the R-NS flows
$\DD(\uu)$ is a constant while it fluctuates in the I-NS flows.

Assuming the ``Chaotic Hypothesis'', \cite{Ru995,Si994,GC995b}, at least
for large $R$ and large $En$ the observable $\a(\uu)$ will reach an average
value, \ie the ``running average'' of $\a$ will
approach a limit, 

\be\frac1t\int_0^t\a(\uu(t))dt\tende{t\to\infty} \int \m^{c}_R(d\uu)
\Eq{e2.3}\ee
(exponentially fast, taking the hypothesis literally).

The proposal is that, {\it although the I-NS and R-NS are very different},
the two equations should be equivalent if the interest is concentrated on
the statistical fluctuations of ``$K$-local'' or ``large scale''
observables, meaning observables depending on the Fourier components
$\uu_\kk$ with $|\kk|<K$ if $K$ is fixed and $R$ or $En$ are large enough
(see, however, Sec.5 below).

More precisely the NS equations are imagined ``regularized'' by restricting the
Fourier components to be $|\kk_i|\le N, i=1,2$ and the distributions
$\m^c_R,\m^{mc}_{En}$ will be imagined to be the limits of the
corresponding distributions for the ``regularized'' equations. Then
\*

\0{\bf Conjecture:} {\it The distributions $\m^c_R\in \EE^c$ and $\m^{mc}_{En}$
will assign, for $R,En$ large enough, the same probability
distribution to $K$-local, time reversible, observables if

\be \m^{c}_R(\DD)= En, \qquad {\rm or}\qquad  \m^{mc}_{En}(\a)=\fra1R\Eq{e2.4}
\ee
}

\0{\it Remarks} (1) The second in Eq.\equ{e2.4} shows that the conjecture can be
interpreted as a ``homogeneization property'', \ie in R-NS flows the
fluctuating $\a$ can be replaced by its average for the purpose of studying
the statistics of local observables, thus leading to the fluctuations in
the I-NS flows.\\
\0(2) The R-NS equation is reversible: therefore under the chaotic hypothesis
a version of the fluctuation theorem, \cite{GC995,GC995b}, should hold for
the distributions in $\EE^{mc}$. However the R-NS equation is a pde and the
phase space contraction, \ie the divergence of the equation
$\s(\uu)={\rm div} (\a(\uu)\D\uu)$, formally given by

\be \s(\uu)=\a(\uu)\sum_\kk\kk^2
-2\a(\uu)\frac{\sum_\kk \kk^6
  \gg_{-kk}\cdot\uu_\kk}{\sum_\kk\kk^4|\uu_\kk||^2}
\Eq{e2.5}\ee
is divergent in the limit $N\to\infty$ and the fluctuation relation needs
to be suitably interpreted, see Sec.5.
\\
\0(3) The above conjecture holds in a sense analogous to the equivalence of
descriptions of equilibria by different ensembles, like microcanonical or
canonical ensembles. The superscripts $c$ and $mc$ are used to stress the
analogy with the canonical ensemble (with $R$ playing the role of the
canonical inverse temperature) and the microcanonical ensemble (with $En$
playing the role of microcanonical energy). In the notion of local
observable the cut-off $K$ plays the role of the finite volume $V_0$ inside
a large container $V$ (which has to be considered in the ``thermodynamic
limit $V\to\infty$'') represented, in the present context, by the cut-off
$N$.  \\
\0(4) The analogy with he equivalence between equilibrium ensembles suggests
intrepreting $R$ as the inverse temperature $\DD(\uu)=En$ as the energy and
$\a(\uu)$ as the kinetic energy: and this suggests the relation

\be \m^c_R(\a)=\frac1R+o(\frac1R)\Eq{e2.6}\ee
(where $o(1/R)$ tends to$0$ faster than $1/R)$)
which is a parameterless relation that could be tested.
\\
\0(5) In other words this is an instance of a general proposition ``{\it In
  microscopically reversible (chaotic) systems \alertb{time reversal
    symmetry} cannot be \alert{spontaneously broken}, but \alertb{only
    phenomenologically so}}'',\cite{Ga998}.
\\
\0(6) A simple further remark is that $En$ in Eq.\equ{e2.4} is exactly
equal to $R$ times the $\m^c_R$-average of the {\it local} observable $\bf
g\cdot\bf u$ and the second average in Eq.\equ{e2.4} is exactly equal to
$En$ times the $\m^{mc}_{En}$ of the same local observable.

\def\SEC{Quantum nonequilibrium ensembles}
\section{\SEC}
\label{sec3}
\iniz

Temperature and heat, defined by the special apparata that measure them,
\cite{NS002}, are important physical observables in \alert{\it
  meso-physics} and \alert{\it nano-physics} studies of nonequilibrium
steady states, \cite{MCT993,ACCCEF011,Ga007}.  In simulations the use of
finite thermostat, \cite{GP009a}, as well as the connection with dynamical
systems theory have been useful. Here the discussion will be restricted to
problems in which there are no external forces and the energy exchanges
only involve heat transfer.

The natural model above, see Figure 1, of a \alert{quantum system} $\CC_0$
coupled to \alert{quantum thermostats} $\TT_1,\TT_2,\ldots$, was proposed
in \cite{FV963} where it was studied \alert{only} with infinite thermostats
consisting of free gases.  A similar semiclassical model goes back to
Ehrenfest and has been considered by many authors, see
\cite{MM978,MCT993,ACCCEF011} for recent accounts, with finite thermostats
as in the \alertb{Ehrenfest thermostat} model described as follows.

Let $H$ be operator on $L_2(\CC_0^{3N_0})$
acting on wave functions $\Ps$ (symmetric or antisymmetric functions of $\V
X_0$),

\be H= -\fra{\hbar^2}2\D_{\V X_0}+ U_0(\V X_0)+\sum_{j>0}\big(U_{0j}(\V
X_0,\V X_j)+U_j(\V X_j)+K_j\big)\Eq{e3.1}\ee
where $\V X_1,\V X_2,\ldots,\V X_n$ are configurations of $N_1,\ldots,N_n$
particles
located in {\it finite} regions $\TT_1,\ldots,\TT_n$ external to $\CC_0$
with densities $\d_j=\frac{N_j}{|\TT_j|}$.

The equations of motion of the system in
$\CC_0\cup\TT_1\cup\ldots\cup\TT_n$ are
\be\eqalign{ &(1)\ -i\hbar {\dot\Ps(\V X_0)}=\,(H(\{\V X_j\}_{j>0})\Ps)(\V
  X_0),\cr &(2)\ \V{{\ddot X}}_j=-\Big(\dpr_j U_j(\V X_j)+ \media{\dpr_j
    U_j(\cdot,\V X_j)}_\Ps\Big)-\a_j \V{{\dot X}}_j,\qquad j>0\cr
  \cr}\Eq{e3.2}\ee
where
\be \media{\dpr_j U_j(\cdot,\V X_j)}_\Ps\defi
  \bra\Ps\dpr_j U_j(\cdot,\V X_j)\ket\Ps=\int |\Ps(X_0)|^2
\dpr_j U_j(\V X_0,\V X_j)\, dX_0\Eq{e3.3}\ee

The Eq.\equ{e3.2} defines a {\it dynamical system} on the \alert{phase
  space} consisting of the points $x=\Big(\Ps,(\{\V X_j\},\ \{\V{{\dot
    X}}_j\})_{j>0}\Big)$ as soon as the ``thermostat
force'' \alertb{$\a_i=\a_i(\Ps,\V X_i,\dot {\V X_i})$ is specified.}

A possible choice, ``isokinetic thermostats'', of the $\a_j$ is:
\be\a_j\defi\fra{\media{W_j}_\Ps-\dot U_j}{2 K_j}, \qquad W_j\defi
  -\V{{\dot X}}_j\cdot \V\dpr_j U_{0j}(\V X_0,\V X_j)\Eq{e3.4}\ee
and with this choice the evolution keeps $K_j\defi\fra12\V{{\dot
    X}}_j^2\defi\fra32 k_B T_j N_j $ \alert{exact constants}. 
\*

\0{\it Remarks:} (1) The above Ehrenfest's dynamics in not a time dependent
Schr\"odinger equation, \cite{MCT993}, and it can be interpreted as the
evolution of a quantum system interacting with $n$ thermostats
$\TT_1,\ldots$ at temperatures $T_1,\ldots$.
\\
(2) The divergence, \ie the dissipation, of the above equations can be
computed directly from the equations of motion Eq.\equ{e3.2}-\equ{e3.4};
it is
\be \s(x)=-\sum_j \Big(\fra{Q_j}{k_B T_j}+ \fra{\dot U_j}{k_B
  T_j}\Big)=-\Big(\sum_j \fra{Q_j}{k_BT_j}\Big) -\frac{d}{dt} \Big(
\sum_j\fra{U_j}{k_B T_j}\Big) \Eq{e3.5}\ee
where $Q_j$ is the work per unit time performed by the test system
particles (located at $\V X_0$) on the particles of the $j$-th thermostat
while the last term in Eq.\equ{e3.5} is a total derivative (work internal
to the $j$-th thermostat) and, therefore, does not contribute to the long
time averages of $\s(x(t))$.  \\
(3) Recalling that there are no nonconservative forces acting on the
system, so that its evolution physically just corresponds to a heat exchange
process, the $-\s(x)$ is naturally interpreted as entropy creation rate
  \\
(4) {\it Time reversal}, \ie change in sign of the velocities $\dot {\V
    X_j}$ and conjugation of the wave function $\Ps(\V X_0)$ is a {\it
    symmetry} for the above dynamics. Hence the equations are reversible
  and (expected to be) chaotic: then accepting the chaotic hypothesis it
  will follow that the stationary state $\m^r_{T_1,\ldots,T_n(dx)}$ will be
  an SRB state and the fluctuations of $\s(x(t))$ satisfy the (stationary
  state) fluctuation relation, \cite{GC995b}.  \*

As the external temperatures $T_1,\ldots,T_n$ vary (keeping the densities
$\d_j$ fixed, for simplicity) the stationary states
$\m^r_{T_1,\ldots,T_n}(dx)$ form an ensemble $\EE^r$ of probability
distributions. The point of view leading to the conjecture discussed in the
previous section about the irrelevance of the particular thermostat used
can be applied here as well.

Therefore it is possible to consider several other thermostat models: for
instance suppose that the external containers $\TT_1,\ldots,\TT_n$ are
infinite containers and that the stationary state
$\m^{\infty}_{T_1,\ldots,T_n}$
  is the stationary state reached starting from an initial state which in
  each container $\TT_j$ is a typical configuration of a Gibbs' state with
  density $\d_j$ and temperature $T_j$. As $T_j$ vary the distributions 
$\m^{\infty}_{T_1,\ldots,T_n}$ form a family which can be denoted
  $\EE^\infty$.  
It is then natural to conjecture:
\*
\0{\bf Conjecture 2:} {\it Given the temperatures $T_1,\ldots,T_N$,
  $\m^r_{T_1,\ldots,T_n}(dx)\in\EE^r$ and
  $\m^\infty_{T_1,\ldots,T_n}(dx)\in\EE^\infty$ assign the same probability
  distribution to localized observables in the limit in which the volume of
  $\CC_0$ grows (homothetically) to $\infty$, \ie in the thermodynamic
  limit.}  \*

If the system is in contact with a single thermostat at temperature $T_1$
it should be true, by consistency, that the local properties of the
stationary distribution should be the same as those of the Gibbs
distribution at temperature $T_1$ of the system enclosed in $\CC_0$ with no
external thermostat.

Even the latter simple property is not necessarily true and it would be
interesting to test it, to provide some support to the more general
conjecture above, \cite{Ga007}.

Other ensembles have been introduced via suitable modifications of the
Erhenfest dynamics: to some of them the conjecture can be adapted, although
not always. For instance not to the studies in
\cite{CP985,MCT993}: the main difference is that the above analysis
requires, in the quantum cases, a sharp separation between thermostats and
test system, with the consequent dissipation taking place as a {\it boundary
  effect} whose particular properties become irrelevant in the
thermodynamic limit (\ie when the ``test system'', in the sense of
\cite{FV963}, becomes large).

It is possible to extend the conjecture 2 to the ensembles considered in
the references just quoted: however the role of the thermodynamic limit is
important; and it is in this limit that conjecture 2 should hold {\it
  exactly}.\footnote{\tiny In the above references, among other remarks,
  emerges the importance of the connection of some of the ensembles with
  the adiabatic approximation (which means that the classical motion of the
  thermostat particles evolves on a time scale much slower than the quantum
  evolution of the system). And it is stressed that such approximation is
  usually not correct, for instance not in the case of the Ehrenfest
  dynamics: on the other hand the conjecture 2 should hold even for the
  Ehrenfest dynamics, provided the thermodynamic limit is taken. In other
  words the difficulty in obtaining agreement between the three kinds of
  thermostats might be due to the ``test system'', in the sense of
  \cite{FV963}, small size.}

The appropriate formulation of the Ehrenfest dynamics has been proposed in
\cite{ACCCEF011}, see \cite{St966} for purely quantum evolutions. The
extension considered is that the Ehrenfest dynamics is a Hamiltonian
dynamics with Hamiltonian
\be \eqalign{
H_E=& \sum_{j=1}^n \frac1{2M_j}\V P_j^2+\int
d \V X_0\lis\Ps(\V X_0)\Big(
-\fra{\hbar^2}{2m}\D_{\V X_0}\cr
&+ U_0(\V X_0)+\sum_{j>0}\big(U_{0j}(\V
X_0,\V X_j)+U_j(\V X_j)\big)\Big)\Ps(\V X_0)
\cr}\ee
where $(\V P_j,\V X_j)$ are the canonical coordinates of the classical
particles (of masses $M_j$) while the other canonical coordinates are the
real and imaginary parts of the components $p_s+iq_s,\,s=0,1,\ldots$ of
$\Ps(\V X_0)$ on a orthonormal basis (arbitrarily fixed).

The Authors refrain from investigating, in systems like quantum electrons
$+$ classical ions, how close the Ehrenfest dynamics is to the
corresponding quantum electrons interacting with quantum ions. But if their
formulation is applied to quantum systems in interaction with {\it
  external} classical systems (as in Sec.3) it leads naturally to the
statement of conjecture 2 on independence of the particular model adopted
for the external thermostats once the thermodynamic limit is added.

Furhermore the proposal in \cite[Sec.5]{ACCCEF011} remarkably provides
stationary states for the Ehrenfest dynamics of an isolated system (given by
the equidistribution on the energy surface of the Hamiltonian $H_E$).

In nonequilibrium cases the difficulty of proving boundedness
of the region of phase space visited, so that a SRB distribution could be
really defined, is still an open problem even for a system in contact with a
single thermostat%
\footnote{\tiny A difficulty already present in the case of purely
  classical systems, \cite{GP010a,GP010b}.}
except, perhaps, when very special thermostat models are assumed, in which
the thermostat forces acts on all particles as in the classical case
discussed in \cite{EM990}, and not just through the boundaries. For
instance in \cite[Sec.7]{ACCCEF011} the case of an Ehrenfest dynamics with
a single temperature Nos\'e--Hoover thermostat is treated proving the
equivalence of the thermostatted stationary state with the canonical
distribution for $H_E$ (at the same temperature).\footnote{\tiny The
  analogous case, \cite{EM990}, of a isokinetic thermostat can be possibly
  treated by the same method thus extending the result to Ehrenfest dynamics.}

There are a few cases in which an exact solution for a problem of quantum
nonequilibrium is known, \cite{ABGM972,JP002,LLMM017}.

Simulations to test the conjecture seem possible and might be attempted in
the future.

\def\SEC{Reversible-irreversible equivalence for NS}
\section{\SEC}
\label{sec4}
\iniz

Here some results that can be derived, so far, in the case of the I-NS e
R-NS equations, continuing the series of simulations in \cite{Ga017a}, will
be briefly presented. They are derived as a first attempt at a systematic
study of the equivalence: {\it although preliminary} they are encouraging
they will require, if confirmed, further analysis.

Several checks of the conjecture have been attempted, beginning with
\cite{GRS004}. In the check it is important to determine first the average
value of the enstrophy $En$ in the I-RS: the average turns out to be
approached quite slowly and this seems to be the main difficulty.

 In the experiments (\ie simulations) the I-RS and R-NS equations had to be
 truncated in momentum space and only with very moderate size $N$ were
 considered (keep in mind that $N\to\infty$ should be taken). It is
 interesting to present a few recent (preliminary) results.  The main
 difficulty is always to determine, fixed the Reynolds number $R$, which is
 the corresponding average enstrophy $En$ in the I-NS flows.

The following picture shows, in a structured ($960$ Fourier modes) and
turbulent ($R=2048$) flow, the {\it slow approach amid strong fluctuations}
of the instantaneous enstrophy to its running average:

\vglue-9.3cm
\hglue1.5cm%
\includegraphics[width=576pt, height=408pt]{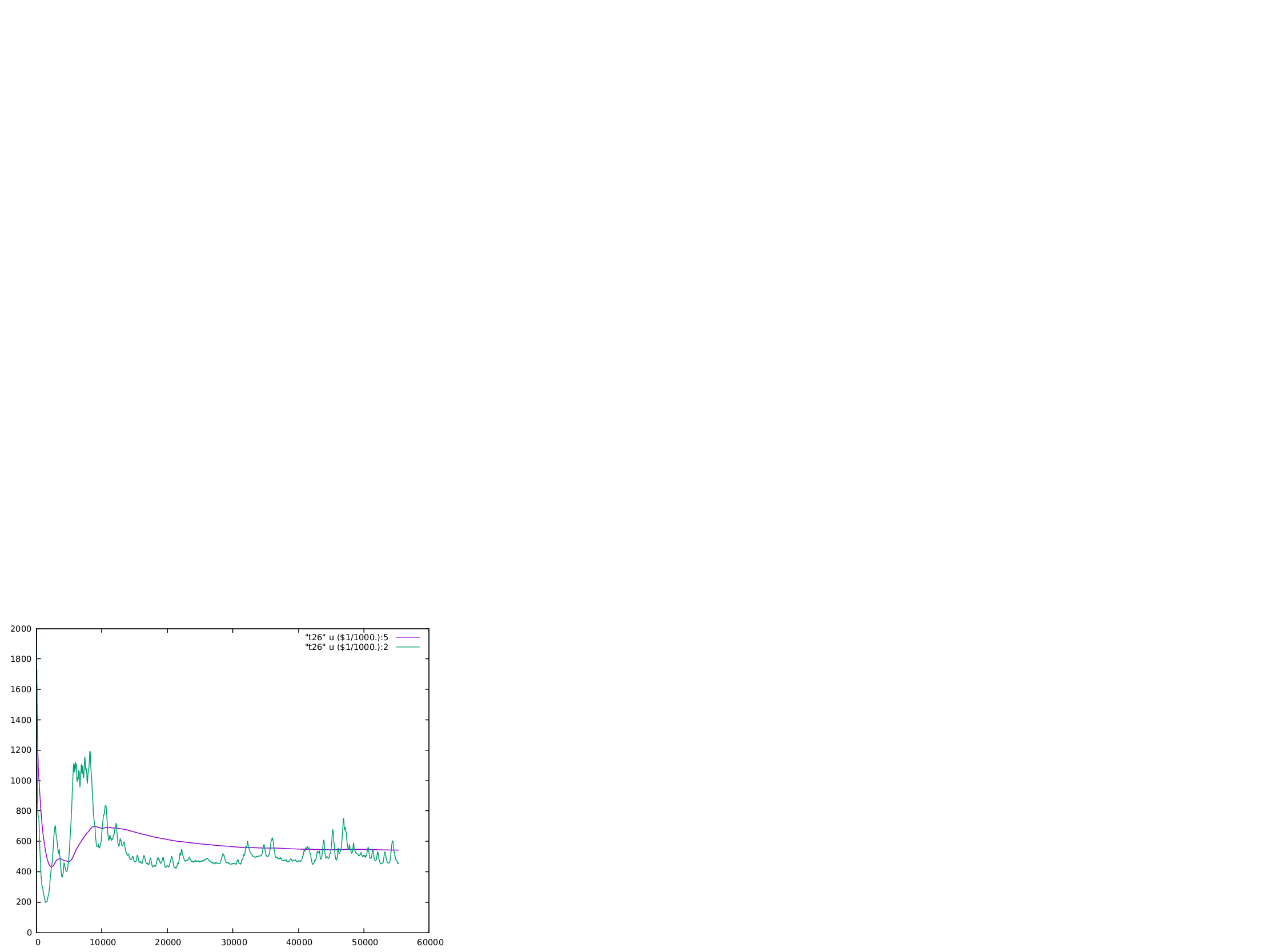}

\begin{spacing}{0.2}\tiny
\0Fig.2: At $960$ modes ($480$ independent complex components), $R=2048$:
the approach and fluctuations of enstrophy; integration step is $h=2^{-15}$
so the time in abscissa is $~1500$ time units (\ie $~50\,10^6$ time
steps). FigEn-64-26-26-4-15\_11\footnote{\tiny The {\it general} coding notation
  is in this case: 64 refers to the cut off size, $N=64/4-1$, 26 says that the
  number of iterations cannot exceed $2^{26}$ in computing the Lyapunov
  exponents (not used here), the second 26 says that the number of
  iterations in computing the running averages cannot exceed $2^{26}$, the
  4 says that is the local Lyapunov exponents with $T_L=2^4$, the 15 mens
  that the integration step is $2^{-15}$ and the 11 indicates that
  $R=2^{11}$.}  \par\end{spacing}

\vglue6mm The simplest check is to test the parameterless relation,
Eq.\equ{e2.6}, between the running average of the reversible viscosity
$\a(\uu)$ in the I-NS flow approaches $\frac1R$ up to $o(\frac1R)$ for
large $R$. And Fig.3 represents for $R=2048$ the evolution under I-NS, the
irreversible NS, of the data $\a(\uu(t))$, their running average and
$1./2048$: the ratio of he two sides of Eq.\equ{e2.5} is $1$ within a few
percent.

\vglue-9.4cm
\hglue1.5cm
\includegraphics[width=576pt, height=408pt]{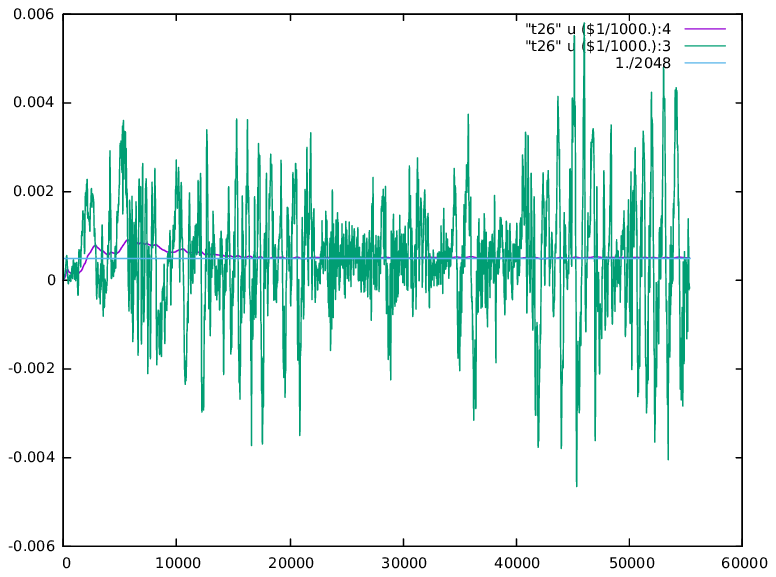}
\*
\begin{spacing}{0.2}\tiny
\0Fig.3: Fluctuations of reversible viscosity $\a(\uu)$ at $960$ modes
($480$ independent complex components), shows coincidence between the
running average of reversible viscosity and the asymptotically expected
value $\frac1R$. The two values, $\fra1R$ and the running average of the
reversible viscosity, agree at the last time within $2.5\%$ but the running
average is still slightly fluctuating, see Fig.3a below (obtained
with a smaller integration step as it seems that
the agreement improves if the integration step is made smaller); but more
accurate study needs to be performed. (FigA2-64-26-26-4-15\_11.0)
\end{spacing}

\vglue-5cm
\hglue1.5cm
\includegraphics[width=360pt, height=255pt]{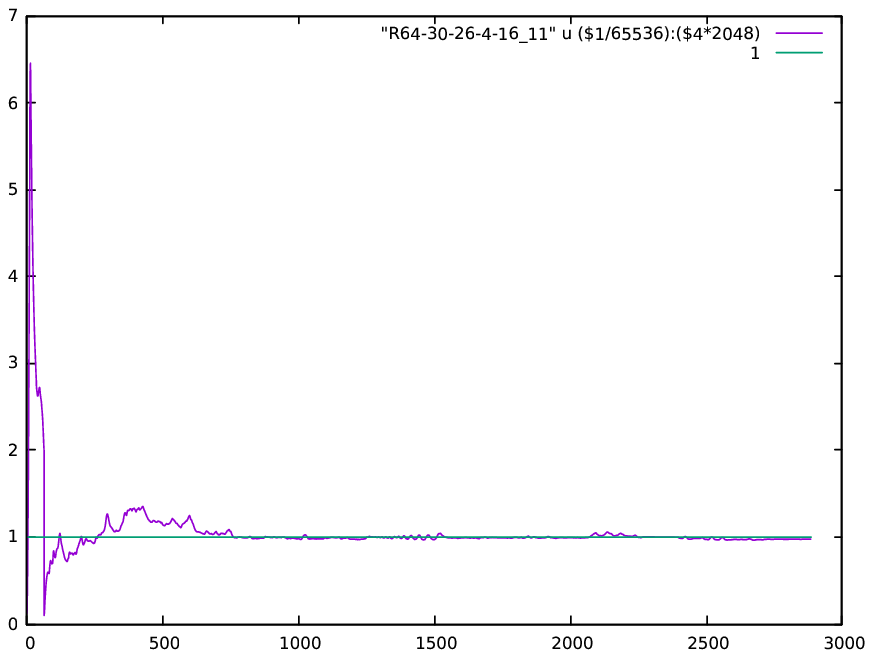}
\vglue3mm
\begin{spacing}{0.2}\tiny
\0Fig.3a: This is the graph of $R\m^c_R(\a)$ which according to
Eq.\equ{e2.5} should tend to $1$ as $R\to\infty$. The integration step
is $\frac12$ of the one in Fig.3 and the number of time units (each of
$2^{16}$ integration steps) is $\sim2900$. (FigA-64-30-26-4-16\_11)
\end{spacing}

\def\SEC{How general is the equivalence? Open problems}
\section{\SEC}
\label{sec5}
\iniz

Looking at the above analysis it emerges that the conjecture should hold
quite generally. It should apply to equations that describe
chaotic many particles systems or to systems that model
macroscopic behavior of systems microscopically subject to the time
reversal symmetry, for instance to the equations described in Sec.2. 

Equations $\dot x=h(x)+f(x), x\in R^N$ may admit a symmetry $I$ with the
property $IS_t=S_{-t}I$, like the truncated Euler equations: if a
thermostat force $-\frac1R L(x)$ is introduced but $N$ is kept fixed (so
that the map $I$ loses its fundamental physical meaning of time reversal
symmetry) it is still possible to formulate equivalence statements: however
the only parameter that can be varied (if $N$ is fixed) is $R$ and
(properly formulated) equivalence can be conjectured at least
asymptotically in the limit $R\to\infty$.

The question has been analyzed in few cases, \eg in the
Lorenz96 model: with encouraging results, \cite{GL014}.
\*

\0(1) How important is the observable that is fixed? in the cases of
Sec.2,3 the enstrophy $\DD(\uu)$ and the kinetic energy of the thermostats
have been considered. The role of the artificial thermostat
force is just to provide a mechanism to keep energy away from
building up due to the work of the driving forces. Therefore several
choices should be possible for determining the multiplier to use to replace
the empirical friction coefficients. In the NS equation discussed in Sec.2,
instead of the enstrophy, the energy $\EE(\uu)=\int_0^{2\p}\uu(\xx)^2 d\xx=
\sum_\kk |\uu_\kk|^2$ could be used: obtaining a reversible viscosity of
the form

\be \a_e(\uu)=
\frac{\int_0^{2\p} \gg(\xx)\cdot\uu(\xx) d\xx}
     {\int_0^{2\p}(\V\dpr \uu)^2(\xx) d\xx}=
     \frac{\sum_\kk \gg_\kk\cdot\uu_{-\kk}}{\sum_\kk\kk^2|\uu_\kk|^2}
     \Eq{e5.1}\ee
For this equation, which can be
called R-NSe and which maintains the total energy $\EE(\uu)$ exactly constant, 
\be\eqalign{
&\alertb{
\dot{\uu}+(\V u\cdot \V\dpr)\uu=-\BDpr p+ \gg+\a_e(\uu)\D\uu,}\qquad
\alert{\BDpr \cdot\uu=0}, \qquad \hbox{\rm [R-NSe]}
\cr
}\Eq{e5.2}\ee
equivalence between corresponding PDF's of the I-NS and R-NSe flows should
also be achieved for $R$ large enough.  The value of $R$ beyond which
a given local observable displays the same statistics in two corresponding
PDF's within a prefixed precision might, however, depend on the observable
and on the regularization cut-off.
\footnote{\tiny The pde's, like the NS equations, have to be truncated to
  be studied in simulations: and the truncation will also affect the size
  of $R$ to realize equivalence of the distribution of the fluctuations in
  corresponding PDF's.} As it is the case in equilibrium statistical
mechanics equivalence may possible for certain choices of the observable
kept constant and not for others: the dissipation $\DD(\uu)$ is a natural
observable to consider.\*
\0
(2) In the case of the I-NS and the R-NSe some equivalence tests have been
performed: with mixed results depending on the size of the truncation:
equivalence has been established at large $R$ in equations strongly
truncated, \cite{GRS004}, but it has been reported to fail at higher
truncations, \cite{RM007}, with Fourier modes with $|\kk|\le10$ (hence
$220$ complex modes) and low $R\sim90$. A more detailed check would be
desirable in this case, for instance probing higher values of $R$ and
determine more accurately the averages of $\EE(\uu)$ or $\DD(\uu)$
corresponding to $R$.  \*
\0(3) In the case of the I-NS equations (not truncated) conjecture 1 could
possibly be extended, to state equivalence with the corresponding R-NS (not
truncated) at all $R$ (even if the stationary motion is laminar and there
may be several attractors). This does not remain true if the equations are
truncated at $|k_i|<N$ with $N$ fixed: as it must be done in the
simulations, and suitable consideration of the limits $R\to\infty$ may
become essential.  \*

\0(4) It is also possible to introduce thermostat forces which fix more
than one constant of motion: a very interesting example is in the work
\cite{SJ993} where the 3D NS equation, truncated at the Kolmogorov scale,
is considered at high $R$. There a reversible friction acts so designed to
enforce the value of the energy in {\it each momentum shell} to equal the
value predicted by the OK $\frac53$ law, \cite{Ga002}. The result is that
the statistical properties of large scale observables in such reversible 3D
NS are essentially the same as those in the simulations of the classical 3D
NS equation truncated at the Kolmogorov scale. Actually, accepting the OK
law, the Kolmogorov scale puts a natural truncation cut off (at $\sim
R^{\frac94}$ modes) and therefore the 3D NS equation is a natural arena
where to formulate and test equivalence conjectures).

For instance the conjecture in Sec.2 would imply that the {\it single
constraint} of constant total dissipation $\DD(\uu)$ should give the same
results as those of the mentioned experiment: a check would be interesting.
\*

\0(5) The conjecture also suggests that the fluctuations of the phase space
contraction $\s(\uu)$, given by the divergence of the reversible
dissipation $\s(\uu)= {\rm div}(\a(\uu)\D\uu)$ (which makes sense in any
truncation of the equations), satisfies, in chaotic regimes, a fluctuation
relation. If $\lis\s$ denotes the infinite time average of $\s(\uu)$ then
the variable $p\defi \frac1\t\int_0^\t \frac{\s(\uu(t))}{\lis\s} dt$ should
have a probability density $P_\t(p)$ verifying the property

\be \frac1\t \log\frac{ P_\t(p)}{P_\t(-p)}\,
    {\lower 5pt\hbox{$\sim$}\atop\raise3pt\hbox{$\scriptstyle\t\to\infty$}}\,
    \k\, p\,\lis\s
\Eq{e5.3}\ee
where $\k$ is a parameterless number: under strong chaoticity assumptions
the above relation has been proposed in \cite{BG997} for reversible
systems. It is natural to ask whether the same variable $\s(\uu)$ follows
the same (one parameter, \ie $\k$) relation also in the I-NS.

It has been mentioned that in the case of the R-NS equations the $\s(\uu)$
is only formally defined: however the fluctuation relation is a property of
the distribution of the running average $\frac1t\int_0^t
\frac{\s(\uu(t))}{\lis{\s}}$. And in a cut-off version of the equation (\eg
setting $\uu_\kk=0$ if $|\kk|>N$ for some $N$ as in the tests of Sec.4
above) from the expression of $\s(\uu)$ in R-NS, Eq.\equ{e2.6}, it appears
that in the ratio $\frac{\s(\uu)}{\lis{\s}}$ the divergent factors
$\sum_\kk \kk^2$ {\it cancel} provided $\k\lis\s$ remains finite, so that it
makes sense to check if a fluctuation relation holds.\\ Some checks of this
property have been performed only in the case of the Lorenz96 model,
\cite{GL014}.

\*
\0(6) The very strong fluctuations of the reversible viscosity observed in
the I-NS, see Fig.5 above, and the equivalence conjecture suggest that it
might be of interest to establish a possible relation between the theory of the
NS equation and its studies under stochastic forces: it would be
interesting to study the NS equation with {\it fixed stirring} force $\gg$
but with {\it random viscosity} with {\it average $\frac1R$} and {\it
  distribution satisfying the fluctuation relation} (\eg a simple Gaussian
white process centered at $\frac1R$ and width determined by the fluctuation
relation).

In this context it should be stressed that there is no theorem of existence
and uniqueness of the R-NS equations (unlike the classical viscous NS
equation, \cite{Ga002}) nor of stochastic NS equations with noise in the
reversible viscosity (in the sense just proposed), even though here only 2D
fluids are considered. The detailed work, \cite{KS012}, could suggest how
to approach the problem.
\*

\0(7) Periodic solutions to I-NS exist, as well as different chaotic
stationary states that can be reached from different initial data,
\cite{FT979,Fr983b,FTZ984,FT985,FGN988}, particularly a low Reynolds
numbers: in such cases the conjecture can still be formulated, see (3)
above, by requiring that the extremal stationary states corresponding to a
given $R$ be in correspondence with equivalent extremal states with the
same $En$, just as in the case of equivalent ensembles at phase transitions
in statistical mechanics (cases in which it is also essential to consider
the thermodynamic limit to obtain equivalence).

\vglue5mm \0{\setstretch{.2}\small {\bf Acknowledgements:} \small {This is
    part of a talk at the ``14th Granada Seminar'' on {\it Quantum Systems
      in and out of Equilibrium: fundamentals, dynamics and applications},
    20-23 June, 2017. Some figures are improved with respect to those
    presented in the talk and I am grateful to IHP for allowing the use of
    a cluster and to L. Biferale for giving me the opportunity to use the
    large cluster at the U. of Roma 2. I am indebted to A. Giuliani for
    several comments that led to modify and clarify the conjecture.}

}

\tiny
\bibliographystyle{unsrt}

\end{document}


\vglue6mm In checking the conjecture I-NS $\sim$ R-NS it was discovered
that also the local Lyapunov exponents of a pair of corresponding PDF's
{\it seem to agree} surprisingly well (as Fig.4 shows).

  \hglue2cm%
  \includegraphics[width=173pt, height=122pt]{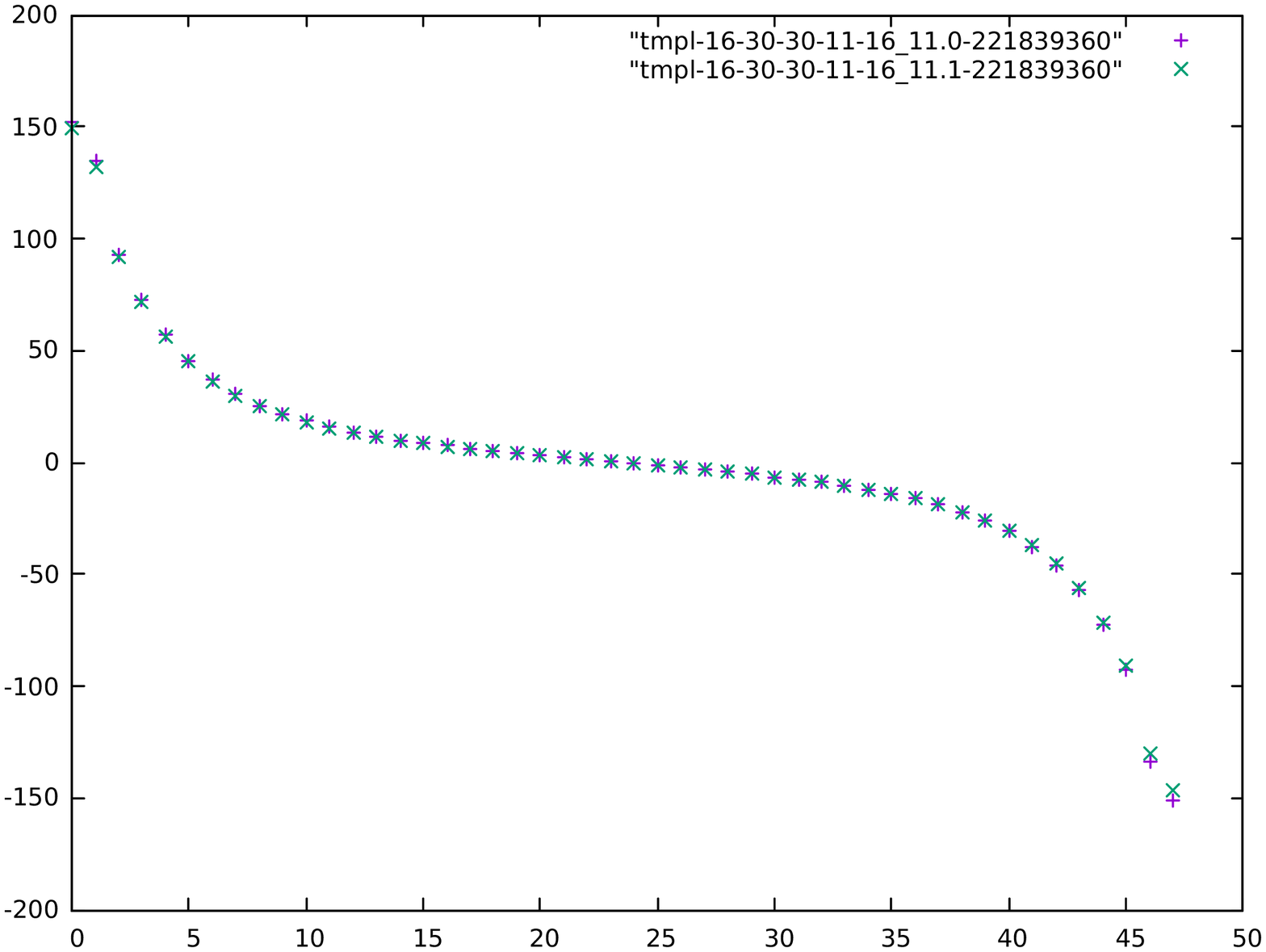}

\begin{spacing}{0.2}\tiny
\0Fig.4: Local Lyapunov spectrum in a $48$ modes truncation ($24$
  independent complex components) of I-NS/R-NS: $(+)$= viscous (I-RS),
  $(\times)$= reversible (R-NS); and $2^{30}$ steps of size 
  $2^{-16}$ (\ie $16384$ time units) and Reynolds number $R=2048$.
  FigL2-16-30-30-16\_11.
\end{spacing}

\vglue6mm This means considering time evolution for a time $T_L$ and
studying the Gram-Schmidt's RU-decomposition of the Jacobian of the
evolution over time $T_L$ and the R-matrix diagonal elements. In fact,
surprisingly, in ref. [30],  the agreement between the two spectra
extended {\it almost} over the entire spectrum. The time $T_L$ chosen for
the local exponents is $2^{11}$ time steps.

Fig.4 is a case with very few Fourier modes.  The next figure Fig.5
represents a case with many more Fourier modes and the same high $R$: in it
there are $500$ time units (\ie each of $2^{16}$ integration steps) and
$T_L$ is $2^7$ integration steps.  The $226$ data are interpolated by lines
serving as eye-guides (if drawn by points the figure is essentially the
same).

\vglue-1cm
\hglue2cm%
\includegraphics[width=173pt, height=122pt]{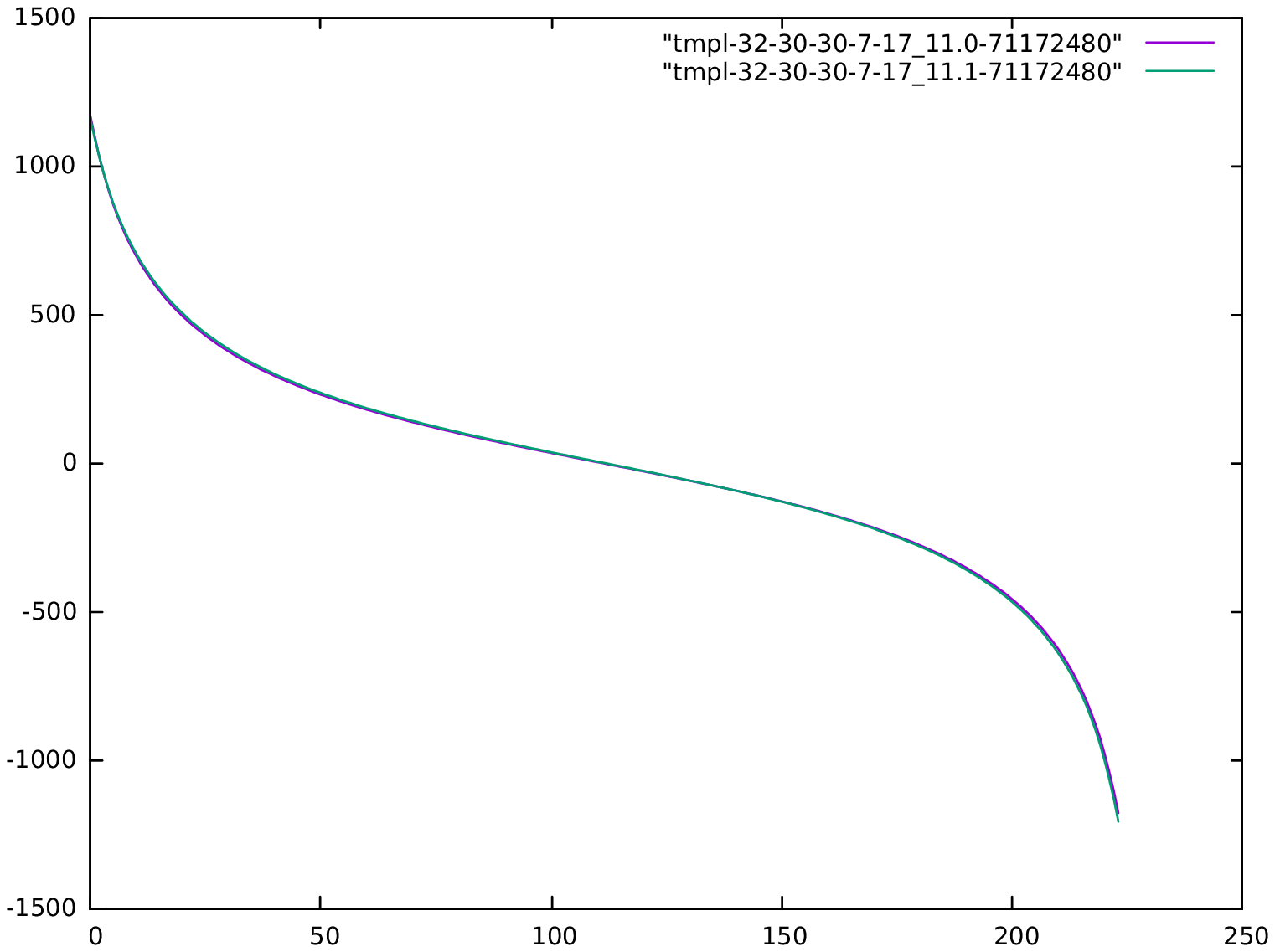}

\begin{spacing}{0.2}\tiny
\0Fig.5: Local Lyapunov spectrum in a $224$ modes truncation ($224$
independent complex components) of I-NS/R-NS. The two spectra, for the I-NS
and R-NS (files ending in $0$ (I-RS), $1$ (R-NS)); and $R=2048$,
$\sim71.e+06$ steps of size $2^{-17}$ (\ie $\sim540$ time units).  The data
are interpolated for visual aid. FigL-32-30-30-7-17\_11.

\end{spacing}

\vglue6mm It is interesting to compare Fig.5 with the following Fig.6:
figure Fig.5 represents the local Lyapunov exponents in both I-NS and R-NS
over $2^{26}$ time steps; while figure Fig.6 below yields the evolution of
the reversible viscosity $\a(\uu)$ in the I-NS flow.

Close examination of the Fig.5 above and Fig.6 below show that, in each
figure, the agreement between the correesponding data represented is within
$\sim5\%$ (it looks better only because of the scale sizes): however it
has to be kept in mind that the regularition is only at $15\times15$ Fourier
modes (and $R$ is ``only'' $2048$):

\vglue-6mm
\hglue2cm%
\includegraphics[width=173pt, height=122pt]{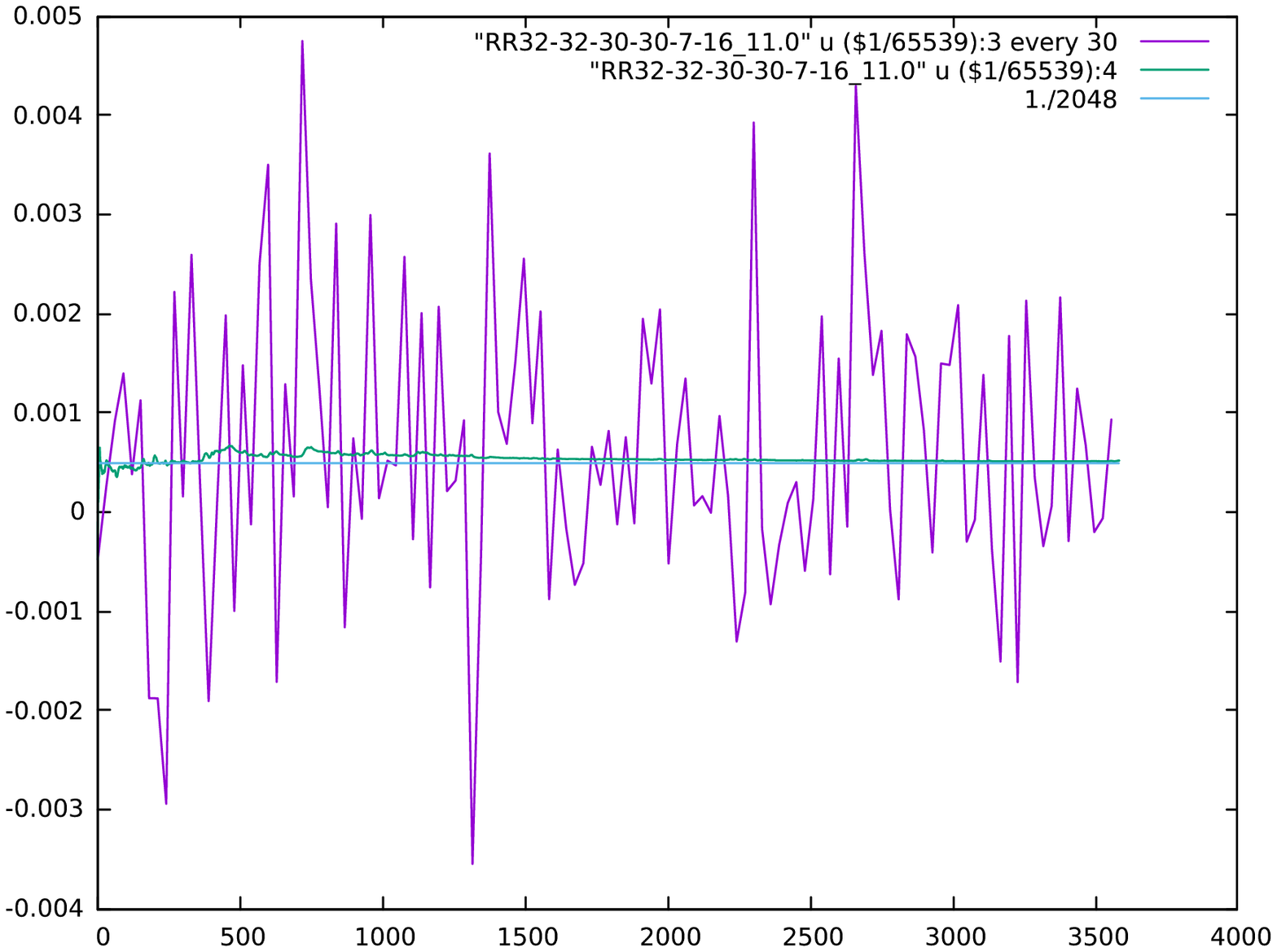}

\*
\begin{spacing}{0.2}\tiny
\0{Fig.6: The data show the evolution of $\a(\uu)$ (skipping, for clarity,
  every $30$ time units)\footnote{\tiny Here $1$ time unit equals $2^{16}$
    integration steps, each of size $2^{-16}$ in the dimensionless units
    used. The data, available now, for Fig.5 are obtained, instead, at a
    finer integrations step, namely $2^{-17}$.} in the I-NS flow observed
  in a run drawn using the same time scale used to produce the Lyapunov
  exponents in Fig.5: hence the fluctuations of $\a$ that took place while
  measuring the local Lyapunov exponents in Fig.5 are only the ones with
  abscissa $<543$).  FigA3-32-30-30-7-16\_11.0.}
  
\end{spacing}

\vglue6mm Therefore, from Fig.6, it is seen that the local exponents
computed in R-NS are driven by a highly fluctuating reversible viscosity
and yet they have essentially the same spectrum as the exponents of the
I-NS, with constant viscosity. The $543$ time units over which the Lyapunov
expondents are averaged to produce Fig.5 has been later extended (so far by
a factor $2$) with the agreement between the exponents for the two flows
improving.